\documentclass[mathleft
]{an}
\usepackage{graphicx}
\usepackage{times}
\newcommand{\vsini}{$v \sin i$}

\newcommand{\logg}{log\,$g$}

\newcommand{\teff}{$T_{\mathrm{eff}}$}

\begin{document}

\Pagespan{1}{}
\Yearpublication{2006}%
\Yearsubmission{2005}%
\Month{11}%
\Volume{999}%
\Issue{88}%

\title{Evidence for weak magnetic fields in early-type emission stars\thanks{Based on observations collected at
ESO, Paranal, Chile (ESO programmes Nos.\ 075.D-0507 and 077.D-0406).}}

\author{S. Hubrig\inst{1}\fnmsep\thanks{Corresponding author:
  \email{shubrig@eso.org}\newline}
\and R.V. Yudin\inst{2,3}
\and M. Pogodin\inst{2,3}
\and M. Sch\"oller\inst{1}
\and G.J. Peters\inst{4}
}

\titlerunning{Magnetic fields in early-type emission stars}
\authorrunning{S. Hubrig et al.}
\institute{
ESO, Casilla 19001, Santiago, Chile
\and
Pulkovo Observatory, Saint-Petersburg, 196140, Russia
\and
Isaac Newton Institute of Chile, Saint-Petersburg Branch, Russia
\and
Space Sciences Center, University of Southern California, University Park, Los Angeles, CA 90089-1341, USA
}

\received{2007}
\accepted{2007}
\publonline{later}


\abstract{
We report the results of our study of magnetic fields in a sample of 15~Be stars using 
spectropolarimetric data obtained at the European Southern
Observatory with the multi-mode instrument FORS\,1 installed at the 8\,m Kueyen telescope.
We detect weak photospheric magnetic fields in
four stars, HD\,56014, HD\,148184, HD\,155806, and HD\,181615.
We note that for HD\,181615 the evolutionary status
is not obvious due to the fact that it is a binary system
currently observed in the initial rapid phase of mass exchange between the two components.
Further, we notify the possible presence of distinct circular polarisation features
in the circumstellar components  of
Ca\,{\sc ii} H\&K in three stars, HD\,58011, HD\,117357, and HD\,181615,
hinting at a probable presence of magnetic fields
in the circumstellar mass loss disks of these stars.
We emphasize the need for future spectropolarimetric observations of Be stars with 
detected magnetic fields to study
the temporal evolution of their magnetic fields and the correlation of magnetic field 
properties with dynamical phenomena taking place in the gaseous circumstellar disks of these stars.
}

\maketitle

\section{Introduction}
\label{sect:intro}

Be stars are defined as rapidly rotating main sequence stars showing normal B-type spectra 
with superposed Balmer emissions.
Further, these stars are characterized by episodic dissipation and formation of a new
circumstellar (CS) disk-like environment, non-radial pulsations and
photometric and spectroscopic variability.
A number of physical processes in classical Be stars (e.g., angular momentum transfer to a 
CS disk, channeling stellar wind matter, accumulation of material 
in an equatorial disk, etc.)
are more easily explainable if magnetic fields are invoked.
In recent years this conclusion was strengthened by numerous theoretical 
works (e.g., Cassinelli et al.\ \cite{2002ApJ...578..951C},
Maheswaran \cite{2003ApJ...592.1156M},
Brown et al.\ \cite{2004MNRAS.352.1061B}).
A detailed discussion concerning the mechanisms involved in formation and evolution 
of disks around Be stars can be found in a recent publication by Owocki (\cite{2006ASPC..355..219O}).

To date we know only three Be stars with magnetic field detections:
$\omega$\,Ori (80$\pm$40\,G -- 
Neiner et al.\ \cite{2003A&A...409..275N}),
$\beta$\,Cep (95$\pm$8\,G --
Donati et al.\ \cite{2001MNRAS.326.1265D}),
and 16\,Peg ($-$156$\pm$31\,G --
Hubrig et al.\ \cite{2006MNRAS.369L..61H}).
These results are cited in the
literature as observational evidence for the existence of magnetic fields 
in Be stars (e.g.\ Brown et al.\ \cite{2004MNRAS.352.1061B}).
However, $\beta$\,Cep has been shown to be a binary star,
where the H$\alpha$ emission line and the magnetic field originate in two different components
(e.g.\ Schnerr et al.\ \cite{2006A&A...459L..21S}),
and the magnetic 
field measured in $\omega$\,Ori is below a 3$\sigma$ threshold.
A longitudinal magnetic field at a level larger than 3$\sigma$ has been
diagnosed for the Be star 16\,Peg (Hubrig et al.\ \cite{2006MNRAS.369L..61H}).
This star has \vsini{}=104~km/s and was classified as a Be star by
Merrill \& Burwell (\cite{1943ApJ....98..153M}) due to the detection of double emission in H$\alpha$.
However, the emission was not confirmed by subsequent observations, and so
the question of the presence of magnetic fields in classical Be stars 
remains open.

\begin{table*}
  \begin{center}
  \caption{The sample of stars observed and the resulting measurements of magnetic fields.}
  \label{tab:kd}
  \begin{tabular}{rcccccccrc}
\hline
\hline
\multicolumn{1}{c}{HD}&
Other &
V &
MJD &
Sp.\ Type &
$v\,\sin\,i$ & 
\teff{} &
\logg {} &
\multicolumn{1}{c}{$\langle$$B_z$$\rangle$} &
Circumstellar \\
 &
identifier &
 &
 &
 &
 [km/s] &
 [K] &
 &
\multicolumn{1}{c}{[G]} &
 Zeeman features \\
\hline
           53367  & V750 Mon       & 7.0 & 53503.496 & B0IVe     & 86  & 28000 & 3.9 & $-9\pm$30           & --- \\
           56014  & EW CMa         & 4.7 & 53512.469 & B3IIIp+sh & 144 & 20000 & 3.8 & {\bf $-$146$\pm$32} & --- \\
           56139  & $\omega$ CMa   & 4.0 & 53502.547 & B2.5Ve    & 109 & 22000 & 4.0 & $-31\pm$24          & --- \\
           57219  & NW Pup         & 5.1 & 53630.820 & B2Vne     & 117 & 23000 & 4.0 & $+80\pm$32          & --- \\
           58011  & NN CMa         & 7.2 & 53502.529 & B1Vnep    & 18  & 25000 & 3.9 & $+135\pm$48         & H+K \\
           58050  & OT Gem         & 6.4 & 53629.914 & B2Ve      & 123 & 23000 & 4.0 & $-101\pm$35         & --- \\
           79351  & a Car          & 3.4 & 53454.590 & B2.5V     & 35  & 22000 & 4.0 & $-7\pm$59           & --- \\
           105435 & $\delta$ Cen   & 2.6 & 53474.622 & B2IVne    & 207 & 23000 & 3.9 & $-30\pm$30          & --- \\
           117357 & CD-61 3819     & 9.1 & 53508.491 & B0.5IIIne & 78  & 27000 & 3.9 & $+7\pm$59           & H+K \\
           137432 & HR 5636        & 5.4 & 53531.660 & B4Ve      & 127 & 18000 & 4.1 & $-58\pm$24          & --- \\
           148184 & $\chi$ Oph     & 4.4 & 53531.722 & B1.5Vpe   & 139 & 24000 & 4.0 & {\bf +83$\pm$21}    & --- \\
           148184 & $\chi$ Oph     & 4.4 & 53862.380 & B1.5Vpe   & 139 & 24000 & 4.0 & {\bf +136$\pm$16}    & --- \\
           148259 & OZ Nor         & 7.4 & 53571.601 & B2Ve      & 86  & 23000 & 4.0 & $+116\pm$49         & --- \\
           153261 & V828 Ara       & 6.3 & 53531.750 & B2IVe     & 184 & 23000 & 3.9 & $-33\pm$41          & --- \\
           155806 & V1075 Sco      & 5.6 & 53531.775 & O7.5 IIIe & 116 & 34000 & 3.8 & {\bf $-$115$\pm$37} & ---  \\
           181615 & $\upsilon$ Sgr & 4.6 & 53519.910 & B2Vpe     & 69  & 23000 & 4.0 & {\bf +38$\pm$10}    & H+K \\
       \hline
  \end{tabular}
  \end{center}
\end{table*}

It is quite possible that the difficulty of detecting magnetic fields in Be
stars is quite similar to that discussed for young Herbig Ae/Be
stars with accretion disks by Hubrig et al.\ (\cite{2007A&A...463.1039H}).
However, in contrast 
to Herbig stars showing evidence of magnetically mediated disk accretion, the Be disks 
originate from ejection or decretion of mass from the rapidly rotating B stars.
The observed spectral lines may form over a relatively large volume, and the magnetic field topology is likely
rather complex. The presence of a mixture of photospheric and 
CS magnetic fields could drive the net 
line-of-sight magnetic flux to near zero values. 
Also, even if there are very strong, small scale magnetic fields distributed over the surface of the star,
these could go undetected in our measurements,
as the measured mean longitudinal magnetic field $\langle$$B_z$$\rangle$
is the average over the stellar hemisphere visible at
the time of observation of the component of the magnetic field parallel to the line of sight,
weighted by the local emergent spectral line intensity.

Below, we report our results of the study of magnetic fields in a sample of 15~Be stars 
carried out with the multi-mode instrument FORS\,1 installed at the 8\,m Kueyen telescope at the VLT.

\section{Analysis}

The observations have been carried out in April-September 2005 in service mode. 
Using the narrowest available slit width of 0\farcs4 the achieved spectral resolving power of the FORS\,1 spectra 
obtained with the GRISM 1200g was about 4000 in the spectral region 
$\lambda\lambda$\,3850--5000\,\AA{}.
Each observation consisted of 8--10~subexposures with a typical duration 
of the order of tens of seconds for each subexposure.
One additional observation of the Be star HD\,148194 ($\chi$~Oph) was obtained in May 2006 with the 
same instrument and GRISM 600B using the narrowest available slit width of 0\farcs4 to obtain 
a resolving power of about 2000.

A detailed description of the assessment of the longitudinal magnetic field measurements 
using FORS\,1 is presented in our previous papers (e.g., Hubrig et al.\ \cite{2004A&A...415..661H}, 
\cite{2004A&A...415..685H}, and \cite{2004A&A...428L...1H}, and references therein).
The errors of the measurements of the polarization have been determined from photon counting 
statistics and have been converted to errors of field measurements.

\section{Results}

\begin{figure}
\centering
\begin{tabular}{c}
\includegraphics[width=0.42\textwidth]{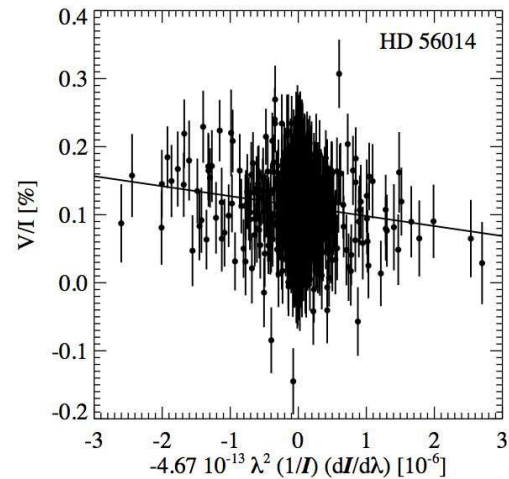}
\end{tabular}
\caption{
Regression detection of the magnetic field in HD\,56014.
}
\label{fig:Fits}
\end{figure}


\begin{figure}
\centering
\includegraphics[width=0.24\textwidth]{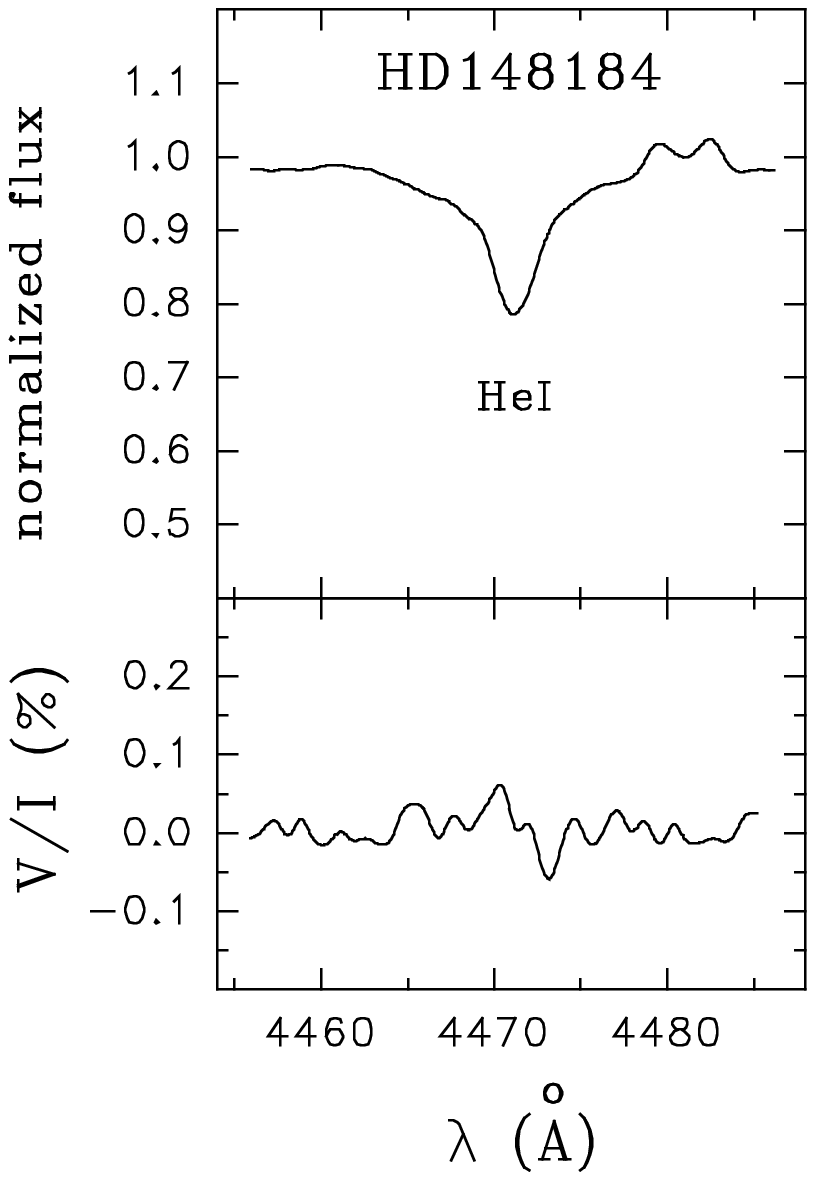}
\includegraphics[width=0.24\textwidth]{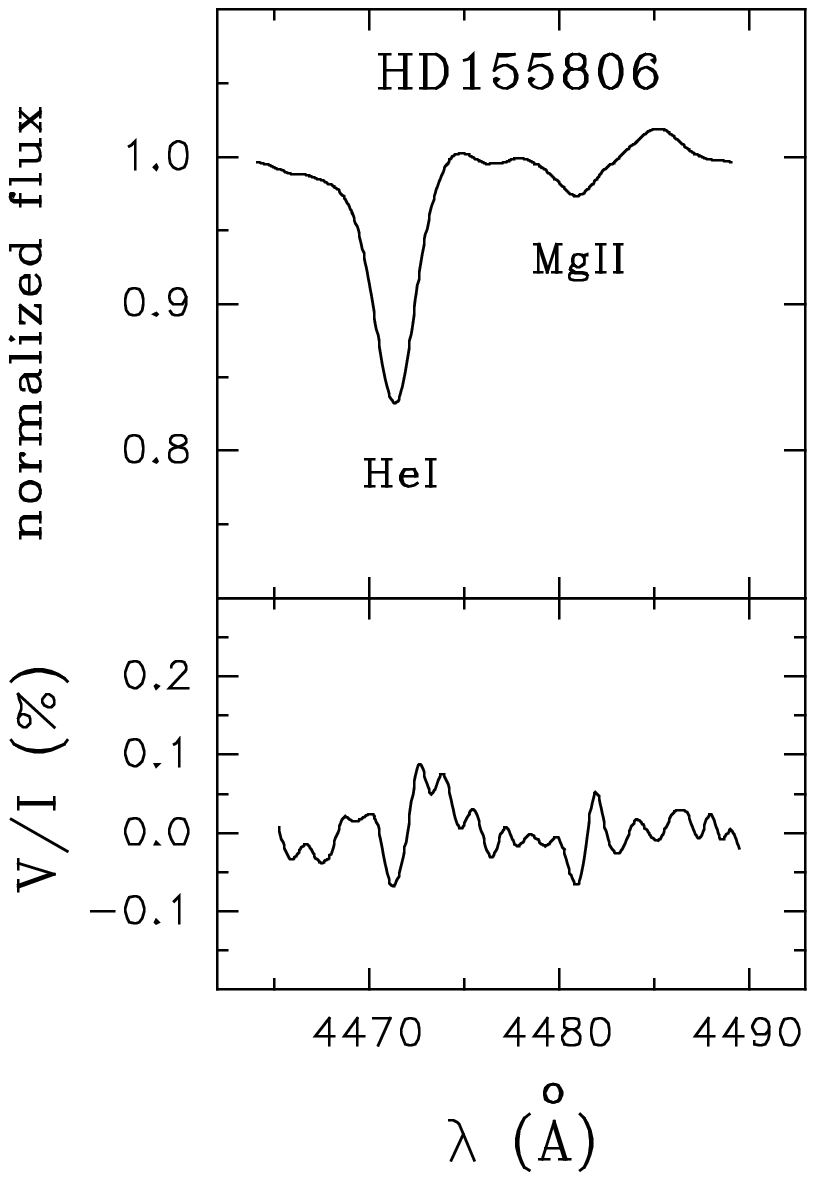}
\caption{Stokes~I and V spectra of HD\,148184 and HD\,155806 in the spectral region around 
the line He\,{\sc i} $\lambda$ 4471.5\,\AA{}. }
\label{fig:IVss}
\end{figure}

The results of our analysis are summarised in Table~\ref{tab:kd}.
In the first four columns we give the HD number of the target, another identifier, the V magnitude, and the
modified Julian date of the observations.
In the next two columns follow the spectral type and the $v\,\sin\,i$ value, which
were both taken from Yudin (\cite{2001A&A...368..912Y}).
In columns seven and eight we present \teff{} and \logg{} values used to calculate the
synthetic spectra, and
in column nine we list the measured mean longitudinal magnetic fields $\langle$$B_z$$\rangle$. 
The mean longitudinal magnetic field is diagnosed from the slope of a linear regression of $V/I$
versus the quantity
$-\frac{g_{\rm eff}e}{4\pi{}m_ec^2} \lambda^2 \frac{1}{I} \frac{{\mathrm d}I}{{\mathrm d}\lambda} \left<B_z\right> + V_0/I_0$.
We show an example of the regression detection for the magnetic field
in HD\,56014 in Fig.~\ref{fig:Fits}.
Our experience from studying a large sample of magnetic and non-magnetic Ap and Bp stars revealed  that this
regression technique is very robust and that detections with $\left<B_z\right> > 3\sigma_z$ result only
for stars possessing magnetic fields.
A longitudinal magnetic field at a level larger than 3$\sigma$ has been detected 
in four stars, HD\,56014, HD\,148184, HD\,155806, and HD\,181615 (also indicated in bold face in Table\,1).
Also, an inspection of the Stokes~V spectra of these four stars reveals noticeable Zeeman 
features at the position of numerous spectral lines.
As an example, we present in Fig.~\ref{fig:IVss} the Stokes~I and V spectra for 
HD\,148184 and HD\,155806 in the spectral region around the line He\,{\sc i} $\lambda$ 4471.5\,\AA{}.

\begin{figure*}
   \centering
  \includegraphics[width=0.32\textwidth]{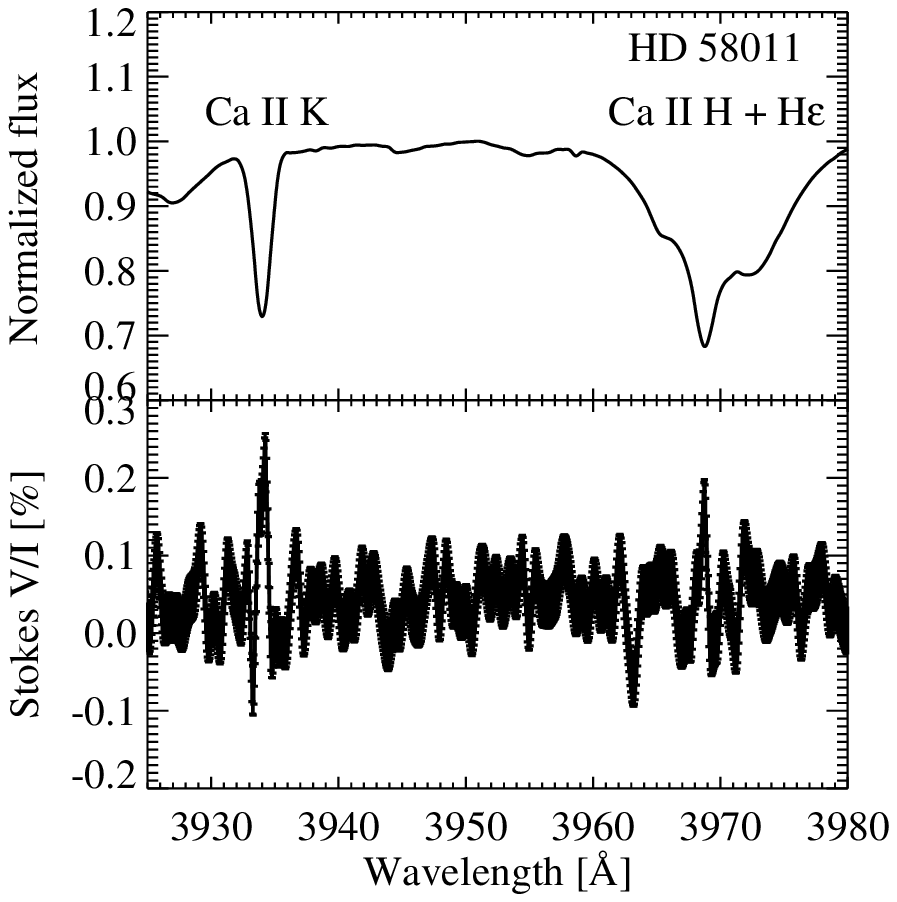}
  \includegraphics[width=0.32\textwidth]{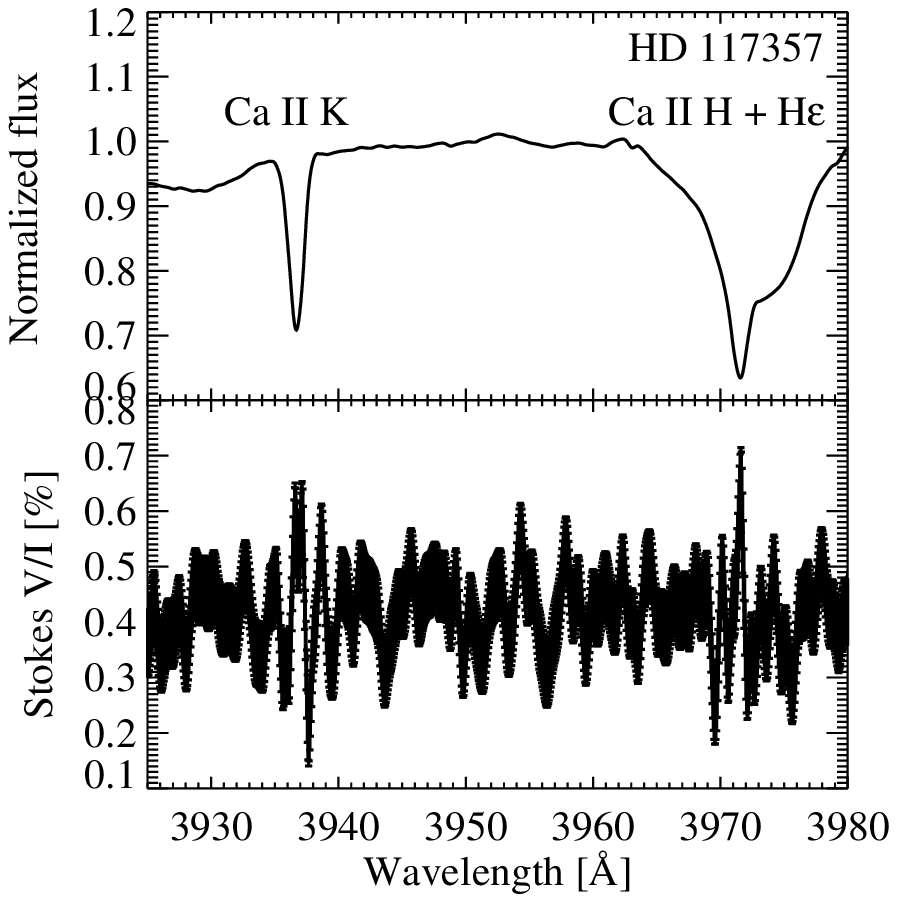}
  \includegraphics[width=0.32\textwidth]{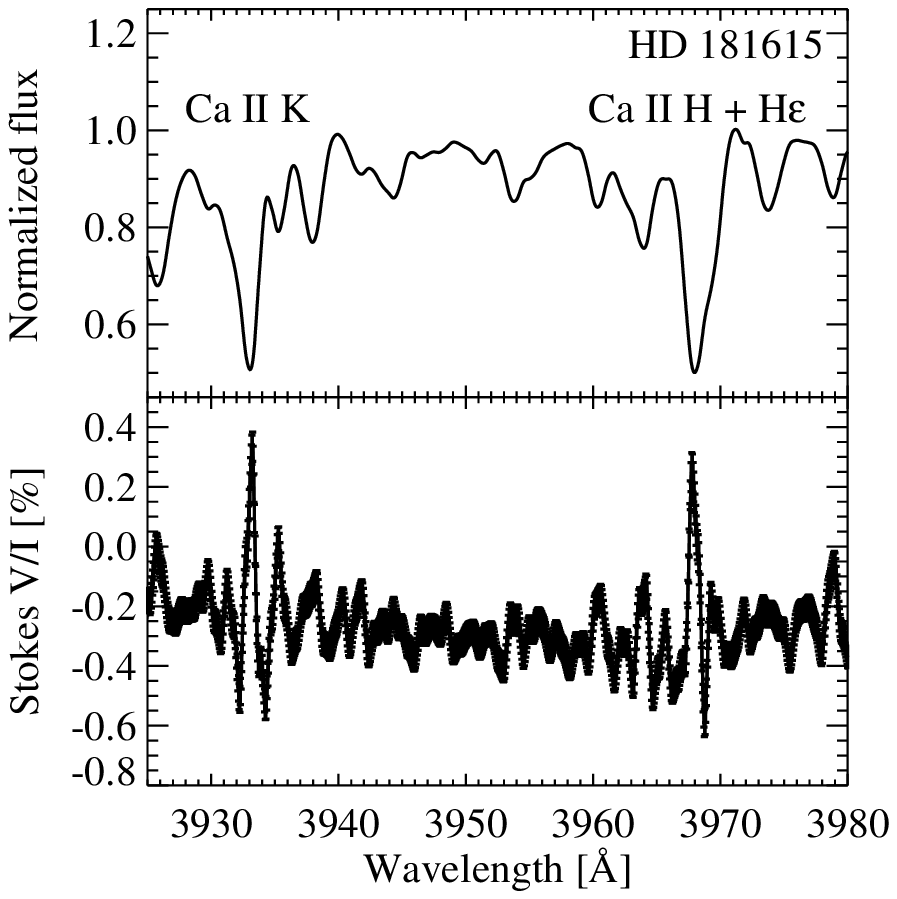}
   \caption{Stokes~I and V spectra of HD\,58011, HD\,117357, and HD\,181615 around the Ca\,{\sc ii} doublet.
The thickness of the plotted lines in the Stokes~V spectra 
corresponds to the uncertainty of the measurement of polarization determined from photon noise.
}
              \label{Fig3}%
    \end{figure*}

According to 
Porter \& Rivinius (\cite{2003PASP..115.1153P}),
the photospheric spectra of Be stars are frequently 
superposed by strong absorption spectra from Be CS disks.
For three stars in our sample, HD\,58011, HD\,117357, and HD\,181615,
we noticed the presence of distinctive circular polarization 
signatures detected in the Stokes~V spectra of the Ca\,{\sc ii} H\&K lines
which appear unresolved at the low spectral resolution achievable with FORS\,1 (R$\sim$4000),
denoted in column~10 of Table~\ref{tab:kd}. 
The profiles of these Ca lines in the FORS\,1 spectra taken in integral light 
are deeper than predicted by
synthetic spectra computed with the  
code SYNTH\,+\,ROTATE developed by Piskunov (\cite{1992stma.conf...92P}).
In Fig.~\ref{Fig3} we present 
Stokes~I spectra (upper panel) and Stokes~V spectra (lower panel)
around the Ca\,{\sc ii} resonance doublet for the stars HD\,58011, HD\,117357, and HD\,181615.
HD\,58011, the Be star with numerous strong emission lines in the visible spectrum, was reported to be 
variable  with an amplitude of 0\fm{}25 after the Hipparcos mission by
Adelman, Mayer \& Rosidivito (\cite{2000IBVS.5008....1A}).
However, the magnetic field 
is determined only at a 2.8$\sigma$ level ($\langle$$B_z$$\rangle$\,=\,+135$\pm$48\,G).
Not much is known about the rather faint Be star HD\,117357 (m$_{\rm V}$=9.1).
Wiegert \& Garrison (\cite{1998JRASC..92..134W}) report about 
the presence of variable emission in Balmer hydrogen lines. In our Stokes~I spectra 
the emission lines are quite weak, and the magnetic field is not detectable.
A detailed discussion on the peculiar spectrum of the system HD\,181615 was presented by
Koubsk{\'y} et al.\ (\cite{2006A&A...459..849K}) who suggested that the visible spectrum
may actually be a combined spectrum of the disk rim and disk face.
As our FORS\,1 spectra are taken at a rather low resolving power, it is 
presently not possible to correctly ascertain the origin of these  Ca\,{\sc ii} H\&K lines.
Additional high-resolution high signal-to-noise spectroscopic observations are needed 
to study the Ca line profiles to be able to decide whether they are formed in the CS disks around 
these two stars.

              \label{Fig5}%



Below we present brief notes on individual objects with detected
photospheric magnetic fields.

\subsection{HD\,56014}

Rivinius, {\v S}tefl \& Baade (\cite{2006A&A...459..137R}) describe this star as a conventional 
Be shell star with narrow absorption lines
and with central quasi emission bumps in photospheric lines 
detected in several Fe\,{\sc ii} lines and in Mg\,{\sc ii} $\lambda$ 4481\,\AA{}.
The magnetic field for this star
is detected at a 4.5$\sigma$ level ($\langle$$B_z$$\rangle$\,=\linebreak[3]\,$-$146$\pm$32\,G).
According to the Washington Double Star Catalog (WDS
-- Worley \& Douglass \cite{1997A&AS..125..523W}) and Mason et al.\ (\cite{1997AJ....114.2112M}) this object
is a visual binary (WDS 07143-2621) with an angular separation of 0\farcs{}150 and an orbital period of about 
180\,yr.

\subsection{HD\,148184}

This Be star with numerous strong emission lines in the Stokes~I spectrum belongs to Upper Scorpius,
which is the youngest of the three subgroups that
form the Scorpius-Centaurus association. 
The variability of the H$\alpha$ line profiles has been reported by
Austin et al.\ (\cite{2004AAS...205.5306A}).
Hubert \& Floquet (\cite{1998A&A...335..565H})
discovered cyclic behaviour of the Hipparcos photometry, but could 
only constrain the period to be $>$0.45\,d.
The magnetic field for this star is detected at a 4$\sigma$ level ($\langle$$B_z$$\rangle$\,=\,+83$\pm$21\,G) 
and at a 8$\sigma$ level ($\langle$$B_z$$\rangle$\,=\,+136$\pm$16\,G).

\subsection{HD\,155806}

This is the hottest star in our sample with spectral type O7.5IIIe 
and is therefore the earliest known star showing emission lines typically 
seen in Be stars (e.g.\ Negueruela, Steele \& Bernabeu \cite{2004AN....325..749N}).
The magnetic field for this star is detected at a level of 3.1$\sigma$ 
($\langle$$B_z$$\rangle$\,=\,$-$115$\pm$37\,G).

\subsection{HD\,181615}

This system is a very rare emission-line binary system of a
strange spectral type and complexity.
The existence of strong H$\alpha$ emission in the visible spectrum was reported by
Campbell (\cite{1895ApJ.....2..177C})  and other investigators.
Very recently, Koubsk{\'y} et al.\ (\cite{2006A&A...459..849K}) concluded that this system with an orbital
period of 138\,d 
is one of very few
known binary systems observed in the initial rapid phase of mass exchange between the two components.
From the photometric and spectroscopic observations Koubsk{\'y} et al.
infer the presence of bipolar jets perpendicular to the orbital plane, similar to
those found for $\beta$\,Lyr, and argue that the peculiar character of the line spectrum of the
brighter component could also be understood as originating from a pseudo-photosphere of
an optically thick disk.
The magnetic field is detected at a level of 3.8$\sigma$
($\langle$$B_z$$\rangle$\,=\,+38$\pm$10\,G).

\section{Discussion}

The detected magnetic fields in the studied Be stars are rather weak, with the largest longitudinal magnetic
field of $-$146\,G measured in the Be star HD\,56014.
These results suggest that strong large scale organized magnetic fields 
are not common among the group of Be type stars.
Further, we noticed the possible presence of  distinct circular polarisation features
in the CS components  of
Ca\,{\sc ii} H\&K in two Be stars, HD\,58011 and HD\,117357, as well as in 
HD\,181615, indicating the possible presence of magnetic fields in the CS mass loss disks.
Interestingly, similar type CS components in Ca\,{\sc ii} H\&K lines have recently been 
discovered by us from FORS\,1 observations of Herbig Ae/Be stars (Hubrig et al.\ \cite{2007A&A...463.1039H}).
The  CS Ca\,{\sc ii} H\&K line profiles in Stokes~I spectra of Herbig Ae/Be stars are frequently quite 
complex and consist of several components which  are usually assumed to be formed at the base of 
the stellar wind and in the accretion gaseous flow. A future careful high resolution high signal-to-noise 
spectropolarimetric study of the temporal behaviour of the Zeeman features in the Stokes~V spectra will 
allow to get the highly desirable insight into the nature of gaseous disks of these two Be stars.

During our observations the exposure times for each target were rather short, of the order of 15--20~min.  
Thus, not much can be inferred with respect to the variability and  evolution of their magnetic fields and 
the correlation of magnetic field properties with dynamical phenomena taking place in these stars. 
To constrain the structure of magnetic fields in Be stars and to probe the presence of localized 
transient magnetic fields suggested by several previous studies (e.g., Mathys \& Smith \cite{2000ASPC..214..316M} and 
references therein), the magnetic field  measurements 
should be carried out over short (minutes) and long (rotational periods) time scales. 
The existence of small scale, i.e.\ highly non-dipole magnetic fields has been suggested by X-ray observations 
of flaring events
(e.g. Smith, Robinson \& Corbet \cite{1998ApJ...503..877S}).

\section*{Acknowledgments}

MP was supported partly by the RSSI grant 07-02-00535.

\end{document}